\documentclass[aps,11pt,twoside]{revtex4}
\usepackage{amsmath,latexsym,amssymb,verbatim,enumerate,graphicx}

\usepackage{hyperref}

\usepackage{theorem}
\newtheorem{definition}{Definition}[section]
\newtheorem{proposition}[definition]{Proposition}
\newtheorem{lemma}[definition]{Lemma}

\newtheorem{theorem}{Theorem}
\newtheorem{corollary}[definition]{Corollary}

\def\squareforqed{\hbox{\rlap{$\sqcap$}$\sqcup$}}
\def\qed{\ifmmode\squareforqed\else{\unskip\nobreak\hfil
\penalty50\hskip1em\null\nobreak\hfil\squareforqed
\parfillskip=0pt\finalhyphendemerits=0\endgraf}\fi}
\def\endenv{\ifmmode\;\else{\unskip\nobreak\hfil
\penalty50\hskip1em\null\nobreak\hfil\;
\parfillskip=0pt\finalhyphendemerits=0\endgraf}\fi}
\newenvironment{proof}[1][Proof]{\noindent \textbf{{#1~} }}{\qed}

\newcommand{\bra}[1]{\langle #1|}
\newcommand{\ket}[1]{|#1\rangle}

\newcommand{\tr}{\text{tr}}

\newcommand{\id}{\mathbb{I}}

\mathchardef\ordinarycolon\mathcode`\:
\mathcode`\:=\string"8000
\def\vcentcolon{\mathrel{\mathop\ordinarycolon}}
\begingroup \catcode`\:=\active
  \lowercase{\endgroup
  \let :\vcentcolon
  }

\newcommand{\nc}{\newcommand}
\nc{\rnc}{\renewcommand} \nc{\beq}{\begin{equation}}
\nc{\eeq}{{\end{equation}}} \nc{\bea}{\begin{eqnarray}}
\nc{\eea}{\end{eqnarray}} \nc{\beqa}{\begin{eqnarray}}
\nc{\eeqa}{\end{eqnarray}} \nc{\lbar}[1]{\overline{#1}}

\nc{\conv}{\operatorname{conv}}
\nc{\smfrac}[2]{\mbox{$\frac{#1}{#2}$}} \nc{\Tr}{\operatorname{Tr}}
\nc{\ox}{\otimes} \nc{\dg}{\dagger} \nc{\dn}{\downarrow}
\nc{\lmax}{\lambda_{\text{max}}}
\nc{\lmin}{\lambda_{\text{min}}}

\nc{\cA}{{\cal A}} \nc{\cB}{{\cal B}} \nc{\cC}{{\cal C}}
\nc{\cD}{{\cal D}} \nc{\cE}{{\cal E}} \nc{\cF}{{\cal F}}
\nc{\cG}{{\cal G}} \nc{\cH}{{\cal H}} \nc{\cI}{{\cal I}}
\nc{\cJ}{{\cal J}} \nc{\cK}{{\cal K}} \nc{\cL}{{\cal L}}
\nc{\cM}{{\cal M}} \nc{\cN}{{\cal N}} \nc{\cO}{{\cal O}}
\nc{\cP}{{\cal P}} \nc{\cQ}{{\cal Q}} \nc{\cR}{{\cal R}} \nc{\cS}{{\cal S}}
\nc{\cT}{{\cal T}} \nc{\cU}{{\cal U}} \nc{\cV}{{\cal V}}
\nc{\cX}{{\cal X}} \nc{\cW}{{\cal W}} \nc{\cZ}{{\cal Z}}

\nc{\CA}{{\cal A}} \nc{\CB}{{\cal B}} \nc{\CC}{{\cal C}}
\nc{\CD}{{\cal D}} \nc{\CE}{{\cal E}} \nc{\CF}{{\cal F}}
\nc{\CG}{{\cal G}} \nc{\CH}{{\cal H}} \nc{\CI}{{\cal I}}
\nc{\CJ}{{\cal J}} \nc{\CK}{{\cal K}} \nc{\CL}{{\cal L}}
\nc{\CM}{{\cal M}} \nc{\CN}{{\cal N}} \nc{\CO}{{\cal O}}
\nc{\CP}{{\cal P}} \nc{\CQ}{{\cal Q}} \nc{\CR}{{\cal R}} \nc{\CS}{{\cal S}}
\nc{\CT}{{\cal T}} \nc{\CU}{{\cal U}} \nc{\CV}{{\cal V}}
\nc{\CX}{{\cal X}} \nc{\CW}{{\cal W}} \nc{\CZ}{{\cal Z}}

\nc{\csupp}{{\operatorname{csupp}}}
\nc{\qsupp}{{\operatorname{qsupp}}} \nc{\var}{\operatorname{var}}
\nc{\rar}{\rightarrow} \nc{\lrar}{\longrightarrow}
\nc{\poly}{\operatorname{poly}}
\nc{\polylog}{\operatorname{polylog}} \nc{\Lip}{\operatorname{Lip}}
\nc{\mb}[1]{\mathbf{#1}}
\nc{\ep}{\epsilon}
\nc{\Om}{\Omega}
\nc{\wt}[1]{\widetilde{#1}}

\def\>{\rangle}
\def\<{\langle}

\nc{\glneq}{{\raisebox{0.6ex}{$>$}  \hspace*{-1.8ex} \raisebox{-0.6ex}{$<$}}}
\nc{\gleq}{{\raisebox{0.6ex}{$\geq$}\hspace*{-1.8ex} \raisebox{-0.6ex}{$\leq$}}}

\nc{\RR}{{{\mathbb R}}}
\nc{\FF}{{{\mathbb F}}}
\nc{\HH}{{{\mathbb H}}}
\nc{\NN}{{{\mathbb N}}}
\nc{\ZZ}{{{\mathbb Z}}}
\nc{\PP}{{{\mathbb P}}}
\nc{\QQ}{{{\mathbb Q}}}
\nc{\UU}{{{\mathbb U}}}
\nc{\WW}{{{\mathbb W}}}
\nc{\EE}{{{\mathbb E}}}
\rnc{\SS}{{{\mathbb S}}}

\nc{\vholder}[1]{\rule{0pt}{#1}}
\nc{\wh}[1]{\widehat{#1}}
\nc{\h}[1]{\widehat{#1}}

\nc{\ob}[1]{#1}

\def\beq{\begin {equation}}
\def\eeq{\end {equation}}

\nc{\eq}[1]{Eq.~(\ref{eq:#1})} \nc{\eqs}[2]{Eqs.~(\ref{eq:#1}) and
(\ref{eq:#2})}

\nc{\eqn}[1]{Eq.~(\ref{eqn:#1})}
\nc{\eqns}[2]{Eqs.~(\ref{eqn:#1}) and (\ref{eqn:#2})}

\nc{\region}{\cS\cW}

\begin{document}

\title{{\Large A Reversible Theory of Entanglement and its Relation 
to the Second Law}}

\author{Fernando G.S.L. Brand\~ao}
\email{fernando.brandao@imperial.ac.uk}
\affiliation{Institute for Mathematical Sciences, Imperial
College London, London SW7 2BW, UK}
\affiliation{QOLS, Blackett Laboratory, Imperial College
London, London SW7 2BW, UK}

\author{Martin B. Plenio}
 \email{m.plenio@imperial.ac.uk}
\affiliation{Institute for Mathematical Sciences, Imperial
College London, London SW7 2BW, UK}
\affiliation{QOLS, Blackett Laboratory, Imperial College
London, London SW7 2BW, UK}


\begin{abstract}

We consider the manipulation of multipartite entangled states 
in the limit of many copies under quantum operations that 
asymptotically cannot generate entanglement. As announced in 
[Brand\~ao and Plenio, Nature Physics \textbf{4}, 8 (2008)], and in stark 
contrast to the manipulation of entanglement under local operations and 
classical communication, the entanglement shared by two or more 
parties can be reversibly interconverted in this setting. The 
unique entanglement measure is identified as the regularized 
relative entropy of entanglement, which is shown to be equal 
to a regularized and smoothed version of the logarithmic 
robustness of entanglement. 

Here we give a rigorous proof of this result, which is fundamentally 
based on a certain recent extension of quantum Stein's Lemma proved in 
[Brand\~ao and Plenio, Commun. Math. \textbf{295}, 791 (2010)], giving the best measurement 
strategy for discriminating several copies of an entangled state 
from an arbitrary sequence of non-entangled states, with an optimal 
distinguishability rate equal to the regularized relative entropy 
of entanglement. We moreover analyse the connection of our approach 
to axiomatic formulations of the second law of thermodynamics.

\end{abstract}

\maketitle

\parskip .75ex


\section{Introduction}

A basic feature of many physical settings is the existence of 
constraints on physical operations and processes that are available. 
These constraints generally imply the existence of resources 
that can be consumed to implement operations that are otherwise
forbidden due to the constraints that have been imposed. Examples 
include an auxiliary heat bath in order to decrease the entropy 
of a thermodynamical system \cite{Cal85} or prior secret 
correlations for the establishment of secret key between two 
parties who can only operate locally and communicate by a public
channel \cite{Mau98}. In quantum information theory one often 
considers the scenario in which two or more distant parties want 
to exchange quantum information, but are restricted to act locally 
on their quantum systems and communicate classical bits only. A 
resource of intrinsic quantum character, \textit{entanglement}, 
allows the parties to completely overcome the limitations caused 
by the locality requirement on the quantum operations available 
\cite{BBC+93}. 

Resource theories are considered in order to determine when a 
physical system, or a state thereof, contains a given resource; to 
characterize the possible conversions from a state to another when 
one has access only to a restricted class of operations which cannot
create the resource for free; and to quantify the amount of such a 
resource contained in a given system. 

One may try to analyse the above questions at the level of individual 
systems. However, it is natural to expect that a simplified theory 
will emerge when instead one looks at the bulk properties of a 
large number of systems. An illustrative example of such a type of 
theory is thermodynamics, which describes the physics of bulk 
properties of large systems in equilibrium by a very simple set of 
rules of universal character. In the context of its second law, in 
particular, the theory determines in terms of a unique quantity - 
the \textit{entropy} - which transformations from one thermodynamical 
equilibrium state into another are possible by means of adiabatic 
processes. There is a long history of examinations of the 
foundations underlying the second law, starting with Carath\'eodory 
work in the beginning of last century \cite{Car1909}. Of particular 
interest in the present context is the work of Giles \cite{Gil64} 
and notably Lieb and Yngvason \cite{LY99, LY00}, stating that there exists 
a \textit{total ordering} of equilibrium thermodynamical states that 
determines which state transformations are possible by means of an 
adiabatic process. From simple, abstract, axioms one can show the 
existence of an entropy function $S$ fully determining the achievable transformations: given two equilibrium states $A$ and $B$, $A$ can be converted by an adiabatic process 
into $B$ {\em if, and only if,} $S(A) \leq S(B)$. 

It was  noted early on in the development of entanglement theory 
that the same total order for state transformations is found in the 
manipulation of bipartite pure states by local operations and 
classical communication (LOCC), in the asymptotic limit of an
arbitrarily large number of copies of the states. Given two 
bipartite pure states $\ket{\psi_{AB}}$ and $\ket{\phi_{AB}}$, 
the former can be converted into the latter by local operations 
and classical communication (LOCC) if, and only if, $E(\ket{\psi_{AB}})
\geq E(\ket{\phi_{AB}})$, where $E$ is the \textit{entropy of 
entanglement} \cite{BBPS96}, given by the von Neumann entropy of 
either of the two reduced density matrices of the state. 

For mixed bipartite states or pure states of more than two parties, 
however, such a pleasingly simple situation does not hold true 
anymore. There are examples of mixed bipartite states, known as 
\textit{bound} entangled \cite{HHH98}, that require a non-zero 
rate of pure state entanglement for their creation by LOCC in the 
limit of many copies, but from which no pure state entanglement 
can be extracted at all \cite{HHH98, VC01, YHHS05}. As a consequence, 
in the general case for the manipulation of entanglement by LOCC
there is no unique entanglement measure and it is not possible to
establish a direct connection to the axiomatic formulation of the 
second law of thermodynamics. 

In this paper we introduce a class of quantum operations that can 
be considered as the natural counterpart of adiabatic processes in 
entanglement theory, in the sense that it allows us to formulate a 
theory of entanglement manipulation with the same \textit{structural} 
form as the second law of thermodynamics. The main technical tool 
for establishing this result is the generalization of quantum 
Stein's Lemma \cite{HP91, ON00} developed in Ref. \cite{BP08}, 
which allows us to determine the best strategy and the optimal 
distinguishability rate for the discrimination of several copies 
of a given entangled state from an arbitrary sequence of separable 
states.  

\textbf{Structure:} The paper is organized as follows. In section 
\ref{qnslet} we motivate the class of quantum operations that we 
are going to consider for the manipulation of entanglement, while 
in subsection \ref{related} we comment on previous related work. 
In section \ref{defmain5}, in turn, we present a few definitions 
and the main results of the paper. Section \ref{proofmaint5} is 
devoted to the proof of Theorem \ref{maintheorem5} and Corollary 
\ref{corref}. We revisit the choice of the operations employed in 
section \ref{mustcan}. Finally, in section \ref{cafslt} we discuss 
the connection of our framework to works on the foundations of 
the second law of thermodynamics, more specifically to the axiomatic 
approach of Lieb and Yngvason.

The results in this paper were announced and discussed in Ref. \cite{BP08b} and Ref. \cite{Hor08}.

\textbf{Notation:} We let ${\cal H}$ be a finite dimensional Hilbert 
space and ${\cal D}({\cal H})$ the set of density operators acting 
on ${\cal H}$. For two states $\rho, \sigma \in {\cal D}({\cal H})$, 
we define the quantum relative entropy of $\rho$ and $\sigma$ as 
$$S(\rho || \sigma) := \tr(\rho(\log(\rho) - \log(\sigma))).$$ 
Given a Hermitian operator $A$, $||A||_1 = \tr(\sqrt{A^{\cal y}A})$ 
stands for the trace norm of $A$ and $\tr(A)_+$ for the trace of 
the positive part of $A$, i.e. the sum of the positive eigenvalues of $A$. The partial trace of
$\rho \in {\cal D}({\cal H}^{\otimes n})$ with respect to the the 
$j$-th Hilbert space is denoted by $\tr_{j}(\rho)$. Given a 
${\cal M} \subseteq \mathbb{R}^{n}$ we define its associate cone 
by $\text{cone}({\cal M}) := \{ x : x = \lambda y, y \in {\cal M}, 
\lambda \in \mathbb{R}_+  \}$ and its dual cone by ${\cal M}^* := 
\{ x : y^{T}x \geq 0 \hspace{0.1 cm} \forall \hspace{0.05 cm} y 
\in {\cal M} \}$. The Bachmann-Landau notation $g(n) = o(f(n))$ 
stands for $\forall k > 0, \exists n_0 : \forall n > n_0, 
\hspace{0.1 cm} g(n) \leq k f(n)$. Logarithms are taken in the 
base 2. The $K$-dimensional maximally entangled state is denoted 
by $\Phi(K) := \sum_{i=1}^K \sum_{j=1}^K \ket{i, i}\bra{j, j}/K$ 
and we set $\phi_2 := \Phi(2)$.

Given a $k$-partite finite dimensional Hilbert space ${\cal H} := 
{\cal H}_{1} \otimes ... \otimes {\cal H}_{k}$, we say that a state 
$\sigma \in {\cal D}({\cal H})$ is separable if it can be written 
as 
\begin{equation} \label{separable}
        \sigma = \sum_j p_j \sigma_{1, j} \otimes ... \otimes 
        \sigma_{k, j},
\end{equation}
for local states $\sigma_{i, j} \in {\cal D}({\cal H}_i)$ and a 
probability distribution $\{ p_j \}$ \cite{Wer89}. If a state is not separable, we say it is \textit{entangled}. The 
set of separable states over ${\cal H}$ is denoted by 
${\cal S}({\cal H})$, or simply ${\cal S}$ when the Hilbert space 
${\cal H}$ is clear from the context.

\section{Asymptotically Non-Entangling Operations} \label{qnslet}

Studies on the connections of entanglement theory and thermodynamics 
date back to the earlier foundational works on the subject 
\cite{PR97, VP98, HH98, HHH98b, PV98}. There it was noted that the 
basic postulates of quantum mechanics and the definition of entangled 
states imply that \textbf{(a)}  \textit{entanglement cannot be 
created by local operations and classical communication}. It was 
argued that this should be seen as a basic law of quantum information 
processing and can be considered as a weak qualitative analogue of 
the second law of thermodynamics, once we make the identification 
of entanglement with order and of LOCC maps with adiabatic processes. 

Local operations and classical communication are the fundamental 
class of operations to be considered in the distant lab paradigm, 
for which the definition of entanglement emerges most naturally. 
However, in view of principle (a) it is important to note that LOCC 
is not the largest class that cannot generate entanglement out of 
separable states. Consider, for instance, the class of separable 
operations, introduced in Ref. \cite{Rai97}. While it is clear 
that a separable map cannot generate entanglement, it turns out that there are separable 
operations which cannot be implemented by LOCC \cite{BDF+99}.  

Are separable maps the largest class of quantum operations that 
cannot create entanglement? As shown in Ref. \cite{CDKL01}, this 
is indeed the case if we allow the use of ancillas. That is, if 
we require that $\Omega \otimes \id_d$, where $\id_d$ is a identity 
map which is applied to a $d$-dimensional ancilla state, does not 
generate entanglement for an arbitrary $d$, then $\Omega$ must be 
a separable superoperator. However, for the following it will be 
important to note that there is yet a larger class of operations 
for which no entanglement can be generated, if we do not require 
that our class of quantum maps is closed under tensoring with the 
identity as above. 
 
\begin{definition}
Let $\Omega: {\cal D}(\mathbb{C}^{d_1} \otimes ... \otimes 
\mathbb{C}^{d_m}) \rightarrow {\cal D}(\mathbb{C}^{d'_1} \otimes ... 
\otimes \mathbb{C}^{d'_m})$ be a quantum operation. We say that 
$\Omega$ is a separability-preserving or a non-entangling map if for 
every separable state $\sigma \in {\cal D}(\mathbb{C}^{d_1} \otimes 
... \otimes \mathbb{C}^{d_m})$, $\Omega(\sigma)$ is a separable 
state. We denote the class of such maps by $SEPP$.
\end{definition}

From its very definition, $SEPP$ is the largest class of operations 
which cannot create entanglement. An example of a completely positive 
map which is separability-preserving, yet is not a separable operation 
is the swap operator. In fact, the class $SEPP$ is even strictly larger 
than the convex hull of separable operations and the composition of 
separable operations with the swap operator \cite{VHP05}. 

We can formulate a quantitative version of (a), stating that 
\textbf{(b)}  \textit{entanglement cannot be increased by local 
operations and classical communication}. Although (b) is clearly 
stronger than the first version discussed, it is not as fundamental 
as (a), because we must assume there is an underlying way to quantify 
entanglement, something that cannot be done in a completely unambiguous 
manner. Here we will focus on two specific entanglement measures as 
the quantitative notion of entanglement needed for (b). The first 
is the relative entropy of entanglement \cite{VPRK97, VP98}, defined 
as
\begin{equation*} 
        E_R(\rho) := \min_{\sigma \in {\cal S}} S(\rho || \sigma),
\end{equation*}
where ${\cal S}$ is the set of separable states. The second is the 
(global) robustness of entanglement \cite{VT99, HN03}, given by
\begin{equation*}
        R_G(\rho) = \min_{\sigma \in {\cal D}, s \in \mathbb{R}_{+}} 
        \left[s : \frac{\rho + s \sigma}{1 + s} \in {\cal S}\right].
\end{equation*}
We choose these two measures because, using them to quantify 
entanglement, LOCC is again not the largest class of operations for 
which (b) is true. Indeed, non-entangling maps are once more the 
largest such class. 


As we will focus on entanglement manipulation in the limit of 
arbitrarily many copies of the state, we can consider an even larger 
class of maps, which are non-entangling only in the asymptotic limit.
We define this class precisely in section \ref{defmain5}, but here we 
would like to anticipate that this class is formed by sequences of 
maps $\{ \Lambda_n \}_{n \in \mathbb{N}}$ - each acting on $n$ copies 
of the individual multipartite Hilbert space - where each $\Lambda_n$ 
generates at most an $\epsilon_n$ amount of entanglement, and such 
that $\epsilon_n$ goes to zero when $n$ grows.

The motivation for identifying the largest class of operations which
cannot create 
entanglement is that we would like to have a class of operations which 
is as powerful as possible, in order to allow for a simple theory of 
entanglement transformations, but also one which does not trivialize 
the theory, in the sense that every state could converted into another 
and there would be no point to talk about entanglement anymore. In 
this context, the class of non-entangling maps, or asymptotically 
non-entangling maps when we look at the manipulation of many copies 
of the state, emerges as a very suitable choice. 


\subsection{Previous Work and Related Approaches} \label{related}

In Ref. \cite{VK02}, the applicability of Giles axiomatic approach 
\cite{Gil64} to entanglement theory was studied. It was shown that 
for pure state bipartite entanglement the same axioms used in the 
derivation of the second law of thermodynamics hold true. Therefore, 
one can derive the uniqueness of the entropy of entanglement following 
the steps taken by Giles in the derivation of the entropy in 
the context of the second law \cite{Gil64}. One of Giles postulates 
is that if two states $A$ and $B$ are both adiabatic accessible from 
another state $C$, either $A$ is adiabatic accessible to $B$ or 
vice-versa (if not both) \cite{Gil64}. In Ref. \cite{MFV04}, it was 
pointed out that this property does not hold in asymptotic mixed 
state entanglement transformations under LOCC, showing the 
inapplicability of Giles approach in the mixed state scenario. 

Various approaches have been considered to enlarge the class of
operations in a way that could lead to reversibility of entanglement
manipulation under such a set of operations. Two closely related
but different routes have been taken here. 

A first approach was considered in \cite{Rai01, EVWW01, APE03}. 
There, entanglement manipulation was studied under the class of 
operations that maps every state with a positive partial transpose 
(PPT) into another PPT state (including the use of ancillas). It 
was realized in \cite{EVWW01} that every state with a non-positive 
partial transpose becomes distillable under PPT preserving operations. 
This eliminates the phenomenon of bound entanglement in a qualitative
level thereby suggesting the possibility of reversibility in this 
setting. This was taken as a motivation for further studies, e.g.
\cite{APE03}, where it was shown that under PPT maps the antisymmetric 
states of arbitrary dimension can be reversibly interconverted into 
pure state entanglement, clearly showing a nontrivial example of 
mixed state reversibility. Unfortunately, no other example have been 
found so far and, hence, reversibility under the class of PPT 
operations remains as an open question. In the multi-partite
pure state setting PPT preserving operations are not sufficient
to ensure reversibility either \cite{IP05}.

In a second approach one considers every PPT state as a free resource
in an LOCC protocol. Then again, every state with a non-positive 
partial transpose becomes distillable \cite{EVWW01}. However, in 
Ref. \cite{HOH02} it was shown, under some unproven but reasonable 
assumptions, that in this scenario one still has irreversibility.
  
The possibility of having reversible transformations of entangled 
states under enlarged classes of operations was also analysed in 
Ref. \cite{HOH02}. In this work the authors considered the analogy 
\textit{entanglement-energy}, first raised in Refs. 
\cite{HH98, HHH98b, Hor08}, complementary to the \textit{entanglement-entropy} 
analogy \cite{PR97, PV98} considered here (see \cite{Hor08} for a discussion of the results of this paper in this context), to argue that a fully 
thermodynamical theory of entanglement could in principle be 
established even considering the existence of bound entanglement. 
However, under some assumptions on the properties of an entanglement 
measure there defined, it was shown that one is unlikely to 
encounter exactly the setting envisioned. Interestingly, it was 
proven that if one has reversibility under a class of operations 
that includes mixing, then the unique measure of entanglement 
governing state transformations is the regularized relative entropy 
from the set of states which are closed under the class of operations 
allowed. 


\subsubsection{Nice Resources}

There is another line of research which our framework is 
connected to: the quest for identifying the \textit{nice resources} 
of quantum information theory, which allow for a simpler theory over 
the unassisted case. The idea here is not to consider what resources 
are useful from the point of view of information processing, but 
actually the ones that are nice in the sense of leading to a marked 
simplification in the resource theory under consideration.   

The first example of such a nice resource is unlimited entanglement 
between sender and receiver for communication over a noisy quantum 
channel. It has been proven in Refs. \cite{BSS+99, BSS+02} that this 
leads to a remarkably simple formula for the quantum and classical 
capacities (which are actually related by a factor of two), which 
in this case is in single-letter form, meaning that no regularization 
is needed, and a direct generalization of Shannon's capacity formula 
for classical noisy channels.   

A more recent example is the use of symmetric side channels for sending quantum information. By the no cloning theorem \cite{Die82, WZ82} we know that it is not possible to reliably send quantum information through a channel which distributes the information symmetrically between the receiver and the environment. It has been shown in Refs. \cite{SSW06, Smi07} that nonetheless such channels are nice resources, as it is possible to derive a single-letter and convex expressions for the symmetric-side-channel-assisted quantum and private channel capacities. Such an approach has recently lead to a breakthrough in quantum information theory, as it was used by Smith and Yard to show that the quantum channel capacity is not additive \cite{SY08}. 

A third example is of course the use of PPT operations and PPT states 
in entanglement theory, as discussed in the previous section.

\section{Definitions and Main Results} \label{defmain5}

We start with the following definition of maps that generate a small
amount of entanglement. 

\begin{definition}\label{epsilonSEPPreserving}
Let $\Omega: {\cal D}(\mathbb{C}^{d_1} \otimes ... \otimes 
\mathbb{C}^{d_m}) \rightarrow {\cal D}(\mathbb{C}^{d'_1} 
\otimes ... \otimes \mathbb{C}^{d'_m})$ be a quantum operation. We 
say that $\Omega$ is an $\epsilon$-non-entangling (or 
$\epsilon$-separability-preserving) map if for every separable 
state $\sigma \in {\cal D}(\mathbb{C}^{d_1} \otimes ... \otimes 
\mathbb{C}^{d_m})$, 
\begin{equation*}
R_{G}(\Omega(\sigma)) \leq \epsilon.
\end{equation*}
We denote the set of $\epsilon$-non-entangling 
maps by $SEPP(\epsilon)$. 

We then define an asymptotically non-entangling operation as given 
by a sequence of trace-preserving CP maps $\{ \Lambda_n \}_{n \in \mathbb{N}}$, 
$\Lambda_n : {\cal D}((\mathbb{C}^{d_1} \otimes ... \otimes 
\mathbb{C}^{d_m})^{\otimes n}) \rightarrow {\cal D}((\mathbb{C}^{d'_1} 
\otimes ... \otimes \mathbb{C}^{d'_m})^{\otimes n})$, such that each 
$\Lambda_n$ is $\epsilon_n$-non-entangling and $\lim_{n \rightarrow 
\infty}\epsilon_n = 0$. 

\end{definition}

It is worth noting that the use of the global robustness to measure 
the amount of entanglement generated is not arbitrary. The reason 
for this choice will be explained in section \ref{mustcan}.

Having defined the class of maps we are going to use to manipulate 
entanglement, we can define the cost and distillation functions, in 
terms of the optimal rate of conversion from and to, respectively, 
the two qubit maximally entangled state
\begin{equation*}
        \phi_2 = \frac{1}{2}\sum_{i=0}^1\sum_{j=0}^1\ket{i, i}\bra{j, j}.
\end{equation*}

\begin{definition}
The entanglement cost under asymptotically non-entangling maps of a 
state $\rho \in {\cal D}(\mathbb{C}^{d_1}\otimes ... \otimes 
\mathbb{C}^{d_m})$ is given by
\begin{eqnarray} \label{costane}
        E_{C}^{ane}(\rho) := \inf_{ \{ k_n, \epsilon_n \} } \left 
        \{ \limsup_{n \rightarrow \infty} \frac{k_n}{n}  :  
        \lim_{n \rightarrow \infty} \left( \min_{\Lambda_n \in 
        SEPP(\epsilon_n)} || \rho^{\otimes n} - \Lambda_n(\phi_2^{
        \otimes k_n})||_1 \right) = 0, \hspace{0.1 cm}  
        \lim_{n \rightarrow \infty} \epsilon_n = 0  \right \}, 
        \nonumber
\end{eqnarray}
where the infimum is taken over all sequences of integers $\{ k_n \}$ 
and real numbers $\{ \epsilon_n \}$. In the formula above 
$\phi_2^{\otimes k_n}$ stands for $k_n$ copies of a two-dimensional 
maximally entangled state shared by the first two parties and the 
maps $\Lambda_n : {\cal D}((\mathbb{C}^{2}\otimes 
\mathbb{C}^{2})^{\otimes k_n}) \rightarrow {\cal D}((\mathbb{C}^{d_1}
\otimes ... \otimes \mathbb{C}^{d_m})^{\otimes n})$ are 
$\epsilon_n$-non-entangling operations. 
\end{definition}

\begin{definition} \label{distane}
The distillable entanglement under asymptotically non-entangling maps 
of a state $\rho \in {\cal D}(\mathbb{C}^{d_1}\otimes ... \otimes 
\mathbb{C}^{d_m})$ is given by 
\begin{eqnarray}
        E_{D}^{ane}(\rho) := \sup_{\{ k_n, \epsilon_n \} } \left 
        \{ \liminf_{n \rightarrow \infty} \frac{k_n}{n} : 
        \lim_{n \rightarrow \infty} \left( \min_{\Lambda_n \in 
        SEPP(\epsilon_n)} || \Lambda_n(\rho^{\otimes n}) - 
        \phi_2^{\otimes k_n}||_1 \right) = 0, \hspace{0.1 cm} 
        \lim_{n \rightarrow \infty} \epsilon_n = 0  \right \}, 
        \nonumber
\end{eqnarray}
where the infimum is taken over all sequences of integers $\{ k_n \}$ 
and real numbers $\{ \epsilon_n \}$. 
\end{definition}

Note that when we do not specify the state of the other parties we 
mean that their state is trivial. Note 
furthermore that the fact that initially only two parties share 
entanglement is not a problem as the class of operations we employ 
include the swap operation. We are now in the position to state the 
main result of the paper.

\begin{theorem} \label{maintheorem5}
For every multipartite state $\rho \in {\cal D}(\mathbb{C}^{d_1} 
\otimes ... \otimes \mathbb{C}^{d_m})$,
\begin{equation}
        E_C^{ane}(\rho) = E_D^{ane}(\rho) = E_R^{\infty}(\rho) 
        := \lim_{n \rightarrow \infty} \frac{E_R(\rho^{\otimes n})}{n}.
\end{equation}
\end{theorem}

We note that in Ref. \cite{VW01} it was shown that in general 
$E_R(\rho \otimes \rho) < 2 E_R(\rho)$. Therefore the limit in the 
definition of the regularized quantity $E_R^{\infty}$ is necessary. 

We find from Theorem \ref{maintheorem5} that under asymptotically 
non-entangling operations, entanglement can be interconverted 
reversibly. From this we can readily show that in this setting 
there is a total order of entangled states. 

\begin{corollary} \label{corref}
For two multipartite states $\rho \in {\cal D}(\mathbb{C}^{d_1} 
\otimes ... \otimes \mathbb{C}^{d_m})$ and  $\sigma \in 
{\cal D}(\mathbb{C}^{d'_1} \otimes ... \otimes \mathbb{C}^{d'_{m'}})$, 
there is a sequence of quantum operations $\Lambda_n$ such that
\begin{eqnarray}
        \Lambda_n \in SEPP(\epsilon_n),  \hspace{1.5 cm} 
        \lim_{n \rightarrow \infty}\epsilon_n = 0, 
\end{eqnarray}
and 
\begin{eqnarray}\label{sublinmain}      
        \lim_{n \rightarrow \infty} || \Lambda_n(\rho^{\otimes n}) - 
        \sigma^{\otimes n - o(n)} ||_1 = 0
\end{eqnarray}
if, and only if, 
\begin{equation}
        E_R^{\infty}(\rho) \geq E_R^{\infty}(\sigma).
\end{equation}
\end{corollary}

We have also identified the regularized relative entropy of
entanglement $E_R^{\infty}$ as the unique entanglement measure in 
this framework. As shown in Ref. \cite{BP08} and discussed in section 
\ref{proofmaint5}, this measure is related to the optimal rate of 
discrimination from many copies of an entangled state to a separable 
states. Therefore, under asymptotically non-entangling operations the 
amount of entanglement of any multipartite state is completely 
determined by how distinguishable the latter is from a state that 
only contains classical correlations. Furthermore, 
we showed in Corollary III.3 of \cite{BP08} that 
\begin{equation} \label{LG5}
        LG(\rho) := \inf_{\{ \epsilon_n\}} \left \{ \limsup_{n 
        \rightarrow \infty} \frac{1}{n} LR_{G}^{\epsilon_n}(\rho^{
        \otimes n}) : \lim_{n \rightarrow \infty} \epsilon_n = 0 
        \right \} = E_R^{\infty}(\rho), 
\end{equation}
where $LR_G(\rho) := \log(1 + R_G(\rho))$ is the log (global) 
robustness of entanglement \cite{Bra05, Dat08b} and 
\begin{equation*}
        LR_{G}^{\epsilon}(\rho) := \min_{\tilde{\rho} \in 
        B_{\epsilon}(\rho)} LR_{G}(\tilde{\rho}), 
\end{equation*}
with $B_{\epsilon}(\rho) := \{ \tilde{\rho} \in {\cal D}({\cal H}) 
: || \rho- \tilde{\rho} ||_1 \leq \epsilon \}$. Hence, we find 
that the amount of entanglement may equivalently and uniquely be
defined in terms of the robustness of quantum correlations to noise 
in the form of mixing. This observation, in particular Eq. 
(\ref{LG5}), will be important in the proof of Theorem 
\ref{maintheorem5}.  

\section{Proof of Theorem \ref{maintheorem5}} \label{proofmaint5}

As mentioned before, the main technical tool for proving Theorem 
\ref{maintheorem5} is an extension of quantum Stein's Lemma 
\cite{HP91, ON00}, which appeared in Ref. \cite{BP08} as Theorem I. 
Here we state the theorem in the particular case of distinguishing 
a given entangled state from separable states, which is sufficient 
for our purposes. 

\begin{theorem} \label{maintheorem}
\cite{BP08} Let $\rho \in {\cal D}({\cal H})$ be an entangled state.

\textsl{Direct part}: For every $\epsilon > 0$ there exists a sequence of POVMs $\{ A_n, \id - A_n \}_{n \in \mathbb{N}}$ such that
\begin{equation*}
        \lim_{n \rightarrow \infty} \tr((\id - A_n) 
        \rho^{\otimes n}) = 0 
\end{equation*}
and for every $n \in \mathbb{N}$ and every separable state $\omega_n \in {\cal D}({\cal H}^{\otimes n})$,
\begin{equation*}
        - \frac{\log \tr(A_n \omega_n)}{n} + \epsilon \geq 
        E_{R}^{\infty}(\rho).
\end{equation*}

\textsl{Strong converse}: For $\epsilon > 0$ and sequence of POVMs 
$\{ A_n, \id - A_n \}_{n \in \mathbb{N}}$ satisfying 
\begin{equation*}
        - \frac{\log( \tr(A_n \omega_n))}{n} - \epsilon \geq 
        E_{R}^{\infty}(\rho)
\end{equation*}
for every $n \in \mathbb{N}$ and every separable state $\omega_n \in {\cal D}({\cal H}^{\otimes n})$,
\begin{equation*}
        \lim_{n \rightarrow \infty} \tr((\id - A_n) \rho^{\otimes n})
        = 1. 
\end{equation*}
According to Proposition III.1 of [] we can express the statement above as follows.
\begin{equation} \label{compactmain}
        \lim_{n \rightarrow \infty} \min_{\omega_n \in {\cal S}
        ({\cal H}^{\otimes n})} \tr( \rho^{\otimes n} - 2^{yn}
        \omega_n)_+ = 
\begin{cases}
0, & y > E_{R}^{\infty}(\rho),\\
1, & y < E_{R}^{\infty}(\rho).
\end{cases}
\end{equation}
\end{theorem}

From this theorem one can already gain an idea of how we are going 
to construct asymptotically non-entangling maps for the creation and 
distillation processes with a rate matching $E_R^{\infty}(\rho)$. 
For entanglement distillation, we consider a sequence of measure-and-prepare 
quantum operations, which first measure the optimal two-outcome POVM 
from the direct part of Theorem \ref{maintheorem}, subsequently either 
preparing approximately $n E_R^{\infty}(\rho)$ copies of $\phi_2$, 
following the outcome associated to $\id - A_n$ corresponding to $\rho^{\otimes n}$, 
or the separable state orthogonal to the maximally entangled state
for the outcome $A_n$ corresponding to a separable state. A simple 
analysis, performed explicitly in section \ref{distent}, shows that 
this family of maps is indeed asymptotically non-entangling and 
distills $\phi_2$ from $\rho$ with any rate smaller than $E_R^{\infty}(\rho)$. 

For the entanglement cost of $\rho$ in terms of $\phi_2$, we use a 
similar construction. We again perform a two outcome POVM, but now 
to check whether we have $n$ copies of $\phi_2$ or a state orthogonal 
to it. For the case corresponding to a maximally entangled state, we 
then prepare a good approximation $\rho_n$ of approximately $n E_R^{\infty}(\rho)$ 
copies of $\rho$, while in the other case we prepare a state which, 
when mixed with $\rho_n$, has the smallest amount of entanglement 
possible. From the converse part of Theorem \ref{maintheorem} 
(which implies in particular Eq. \ref{LG5} \hspace{0.01 cm}\cite{BP08}), we show 
in section \ref{costnonentsec} that the maps are asymptotically 
non-entangling and create $\rho$ from $\phi_2$ with any 
rate bigger than the regularized relative entropy of entanglement of $\rho$. 

It is intriguing that the strong converse part of Theorem 
\ref{maintheorem} not only implies that distillation with a rate higher 
than $E_R^{\infty}(\rho)$ is impossible, but also that the reverse 
process, the formation of $\rho$ from $\phi_2$, is achievable with 
any such a rate. 

\subsection{The Entanglement Cost under Asymptotically non-Entangling 
Maps} \label{costnonentsec}

We start by showing that the entanglement quantified by the log global 
robustness cannot increase by more than a factor proportional to 
$\log(1 + \epsilon)$ under $\epsilon$-non-entangling maps.

\begin{lemma}\label{epsilonincrease}
If $\Lambda \in SEPP(\epsilon)$, then
\begin{equation} \label{143222}
        LR_G(\Lambda(\rho)) \leq \log(1 + \epsilon) + LR_G(\rho).
\end{equation}
\end{lemma}
\begin{proof}
Let $\pi$ be an optimal state for $\rho$ achieving $R_G(\rho)$
\begin{equation*}
        \rho + R_G(\rho) \pi = (1 + R_G(\rho))\sigma,
\end{equation*}
where $\sigma$ is a separable state. We have that
\begin{equation*}
        \Lambda(\rho) + R_G(\rho) \Lambda(\pi) = (1 + R_G(\rho))
        \Lambda(\sigma),
\end{equation*}
with $R_G(\Lambda(\sigma)) \leq \epsilon$. Setting $Z$ to be a state 
for which $\Lambda(\sigma) + \epsilon Z$ is separable, we find 
\begin{equation*}
        \Lambda(\rho) + R_G(\rho) \Lambda(\pi) + 
        \epsilon(1 + R_G(\rho)) Z = (1 + R_G(\rho))\Lambda(\sigma) + 
        \epsilon(1 + R_G(\rho)) Z \hspace{0.1 cm}\in \text{cone}
        ({\cal S}),
\end{equation*}
from which Eq. (\ref{143222}) follows.
\end{proof}
\vspace{0.2 cm}

\begin{proposition} \label{propcost}
For every multipartite state $\rho \in {\cal D}(\mathbb{C}^{d_1}
\otimes ... \otimes \mathbb{C}^{d_2})$,
\begin{equation}
        E_{C}^{ane}(\rho) = E_{R}^{\infty}(\rho).
\end{equation}
\end{proposition}

\begin{proof}
Let $\Lambda_n \in SEPP(\epsilon_n)$ be an optimal sequence of maps 
for the entanglement cost under asymptotically non-entangling maps, 
i.e.
\begin{equation*}
        \lim_{n \rightarrow \infty} || \Lambda_n(\phi_2^{\otimes k_n}) 
        - \rho^{\otimes n}||_1 = 0,\;\;\;\;\;\;\;\; 
        \lim_{n\rightarrow\infty} \epsilon_n = 0,
\end{equation*}
and
\begin{equation*}
        \limsup_{n \rightarrow \infty} \frac{k_n}{n} = 
        E_{C}^{ane}(\rho).
\end{equation*}
Then, from Lemma \ref{epsilonincrease},
\begin{eqnarray*}
        \frac{1}{n} LR_G(\Lambda_n(\phi_2^{\otimes k_n}))  &\leq&  
        \frac{1}{n} LR_G(\phi_2^{\otimes k_n}) + \frac{1}{n} 
        \log(1 + \epsilon_n) \nonumber \\ &=& \frac{k_n}{n} 
        + \frac{1}{n} \log(1 + \epsilon_n),
\end{eqnarray*}
where the last equality follows from $R_G(\phi_2^{\otimes k_n}) 
= 2^{k_n} - 1$. Hence, as $\lim_{n \rightarrow \infty}\epsilon_n = 0$,
\begin{eqnarray*}
        E_R^{\infty}(\rho) = LG(\rho) &\leq& \limsup_{n \rightarrow \infty} 
        \frac{1}{n}LR_G(\Lambda_n(\phi_2^{\otimes k_n})) \nonumber \\ 
        &\leq& \limsup_{n \rightarrow \infty} \left( \frac{k_n}{n} + 
        \frac{1}{n} \log(1 + \epsilon_n) \right) \nonumber \\ 
        &=& E_C^{ane}(\rho).
\end{eqnarray*}
To show the converse inequality, assume w.l.o.g. that $\rho$ is entangled. We consider maps of the form
\begin{equation*}
        \Lambda_n(A) = \tr(A \Phi(K_n)) \rho_n + \tr(A (\id - 
        \Phi(K_n)))\pi_n,
\end{equation*}
where (i) $\{ \rho_n \}$ is an optimal sequence of approximations 
for $\rho^{\otimes n}$ achieving the infimum in $LG(\rho)$ (note the infimum might not be achievable by any 
sequence $\{ \rho_n \}$. In this case, for every $\mu > 0$ we can 
find a sequence $\{ \rho_n^{\mu} \}$ such that $\lim_{n \rightarrow 
\infty} \frac{LR_G(\rho_n^{\mu})}{n} = LG(\rho) + \mu$, proceed as 
in the case where the infimum can be achieved and let $\mu 
\rightarrow 0$ in the end, obtaining the same results), (ii) 
$\log(K_n) = \lceil \log(1 + R_{G}(\rho_n)) \rceil$, and (iii) 
$\pi_n$ is a state such that
\begin{equation} \label{redone1}
        \frac{\rho_n + (K_n-1) \pi_n}{K_n} \in {\cal S},
\end{equation}
which always exists as $K_n \geq 2^{\log(1 + R_G(\rho_n))} = 
1 + R_G(\rho_n)$. As $\pi_n$ and $\rho_n$ are states, each 
$\Lambda_n$ is completely positive and trace-preserving. 

The next step is to show that each $\Lambda_n$ is a 
$1/(K_n-1)$-separability-preserving map. 
From Eq. (\ref{redone1}) we find
\begin{equation*} 
        \frac{\pi_n + (K_n - 1)^{-1} \rho_n}{1 + (K_n - 1)^{-1}} 
        \in {\cal S},
\end{equation*}
and, thus,
\begin{equation*}
        R_G(\pi_n) \leq \frac{1}{K_n - 1}.
\end{equation*}

From Eq. (\ref{redone1}) we have that
\begin{equation*}
        \Lambda_n(I_b) = \frac{\rho_n + 
        (K_n - 1)\pi_n }{K_n} \in {\cal S},
\end{equation*}
where $I_b$ is the separable isotropic state $I_b=\frac{1}{K}\Phi(K)
+ \frac{\id - \Phi(K)}{K(K+1)}$
at the 
boundary of the separable states set, and
\begin{equation} \label{ppp}
        R_G \left(\Lambda_n\left(\frac{\id - \Phi(K_n)}{K_n^2 - 1}
        \right)\right) = R_G(\pi_n) \leq \frac{1}{K_n - 1}.
\end{equation}
From the form of $\Lambda_n$ we can w.l.o.g. restrict our attention 
to isotropic separable input states. Any such state $I(q)$ can be 
written as
\begin{equation*}
        I(q) = q I_b + (1 - q) \frac{\id - \Phi(K)}{K^2 - 1},  
\end{equation*}
with $0 \leq q \leq 1$. From the convexity of $R_G$,
\begin{equation*}
        R_G(\Lambda_n(I(q))) \leq q R_G(\Lambda_n(I_b)) + (1 - q) 
        R_G\left( \Lambda_n \left(\frac{\id - \Phi(K)}{K^2 - 1} 
        \right) \right) \leq \frac{1}{K_n - 1}, 
\end{equation*}
where we used Eq. (\ref{ppp}) and
\begin{equation*}
        R_G(\Lambda_n(I_b)) = 0.
\end{equation*}
We hence see that indeed $\Lambda_n$ is a 
$1/(K_n - 1)$-separability-preserving map.

In Corollary II.1 of Ref. \cite{BP08}, it was proven that 
$E_R^{\infty}(\rho) > 0$ for every entangled state $\rho$. From 
Eq. (\ref{LG5}) we then find that $LG(\rho) = E_R^{\infty}(\rho) > 0$ 
for every entangled state. Therefore 
\begin{equation*}
        \lim_{n\rightarrow\infty} \frac{1}{K_n - 1} \leq 
        \lim_{n\rightarrow\infty} \frac{1}{R_G(\rho_n)} = 0,
\end{equation*}
where the last equality follows from Eq. (\ref{LG5}). Moreover, as 
\begin{equation*}
        \lim_{n \rightarrow \infty} || \rho^{\otimes n} - 
        \Lambda_n(\Phi(K_n)) ||_1 = \lim_{n \rightarrow \infty} || 
        \rho^{\otimes n} - \rho_n ||_1 = 0,
\end{equation*}
it follows that $\{ \Lambda_n \}$ is an allowed sequence of maps 
for $E_C^{ane}(\rho)$ and, thus,
\begin{eqnarray*}
        E_{C}^{ane}(\rho) &\leq&  \limsup_{n \rightarrow \infty} 
        \frac{1}{n} \log(K_n) \nonumber \\ &=& \limsup_{n \rightarrow 
        \infty} \frac{1}{n} \lceil \log(1 + R_G(\rho_n)) \rceil 
        \nonumber \\ 
        &=& LG(\rho)\\
        &=& E_R^{\infty}(\rho).
\end{eqnarray*}
\end{proof}

\subsection{The Distillable Entanglement under non-Entangling Operations} \label{distent}

Before we turn to the proof of the main proposition of this 
section, we state and prove an auxiliary lemma which will 
be used later on. It can be considered the analogue for non-entangling 
maps of Theorem 3.3 of Ref. \cite{Rai01}, which deals with 
PPT maps. 

\begin{lemma} \label{singfractione}
For every multipartite state $\rho \in {\cal D}(\mathbb{C}^{d_1}
\otimes ... \otimes \mathbb{C}^{d_n})$ the singlet-fraction under 
non-entangling maps,
\begin{equation} \label{singletfractionoriginal}
        F_{sep}(\rho; K) := \max_{\Lambda \in SEPP} \tr(\Phi(K)\Lambda(\rho)),
\end{equation}
where $\Phi(K)$ is a $K$-dimensional maximally entangled state 
shared by the first two parties, satisfies
\begin{equation} \label{sfsep}
        F_{sep}(\rho; K) = \min_{\sigma \in \text{cone}({\cal S})}
        \left[\tr (\rho - \sigma)_+ + \frac{1}{K}\tr(\sigma)\right].
\end{equation}
\end{lemma}

\begin{proof}
Due to the $UU^*$-symmetry of the maximally entangled state and the 
fact that the composition of a $SEPP$ operation with the twirling 
map is again a non-entangling operation, we can w.l.o.g. perform the 
maximization over $SEPP$ maps of the form
\begin{equation*}
        \Lambda(\rho) = \tr(A \rho) \Phi(K) + \tr((\id - A)\rho) 
        \frac{\id - \Phi(K)}{K^2 - 1}.
\end{equation*}
Since $\Lambda$ must be completely positive we have $0 \leq A \leq 
\id$. As $\Lambda(\rho)$ is an isotropic state for every input state 
$\rho$, it is separable iff $\tr(\Lambda(\rho)\Phi(K)) \leq 1/K$ 
\cite{HH99}. Hence, we find that $\Lambda$ is non-entangling iff for 
every separable state $\sigma$,
\begin{equation*}
        \tr(A \sigma) \leq \frac{1}{K}.
\end{equation*}
The singlet fraction is thus given by 
\begin{equation*}
        F_{sep}(\rho; K) = \max_{A} [\tr(A \rho) : 0 \leq A \leq \id, 
        \hspace{0.2 cm} \tr(A \sigma) \leq 1/K, \hspace{0.2 cm} 
        \forall 
        \hspace{0.1 cm} \sigma \in {\cal S}]. 
\end{equation*}
The R.H.S. of this equation is a convex optimization problem 
and we can find its dual formulation. Let us form the Lagrangian 
of the problem, 
\begin{equation*}
        L(\rho, A, X, Y) =  - \tr(A \rho) - \tr(X A) - 
        \tr(Y(\id - A)) - \tr((\id/K - A) Z),
\end{equation*}
where $X, Y \geq 0$ are Lagrange multipliers associated to 
the constraints $0 \leq A \leq \id$, and $Z \in \text{cone}
({\cal S})$ is a Lagrange multiplier (an unnormalized separable 
state) associated to the constraint $\tr(A \sigma) 
\leq 1/K \hspace{0.2 cm} \forall \hspace{0.1 cm} \sigma 
\in {\cal S}$. The dual problem is then given by 
\begin{equation*}
        F_{sep}(\rho; K) = \min_{Y, Z} [\tr(Y) + \frac{1}{K} 
        \tr(Z) : Z \in \text{cone}({\cal S}), \hspace{0.1 cm} 
        Y \geq 0, \hspace{0.1 cm} Y \geq \rho - Z]. 
\end{equation*}
Using that $\tr(A)_{+} = \min_{Y \geq A, \hspace{0.01 cm} Y \geq 0} \tr(Y)$, we then find Eq. (\ref{sfsep}).
\end{proof}
\vspace{0.3 cm}

It turns out that to demonstrate that distillable entanglement 
equals the regularized relative entropy of entanglement we do not 
need to allow any generation of entanglement from the maps. In 
analogy to Definition \ref{distane}, we can define the distillable 
entanglement under non-entangling maps as 
\begin{equation}
        E_{D}^{ne}(\rho) := \sup_{\{ k_n \}} \left \{ 
        \liminf_{n \rightarrow \infty} \frac{k_n}{n}  : 
        \lim_{n \rightarrow \infty} \left( \min_{\Lambda_n \in SEPP} 
        || \Lambda_n(\rho^{\otimes n}) - \phi_2^{\otimes k_n}||_1 \right) = 0 \right \}.
\end{equation}

Using Lemma \ref{singfractione} and Theorem \ref{maintheorem} we 
can easily establish the following proposition. 

\begin{proposition} \label{disne}
For every multipartite entangled state $\rho \in 
{\cal D}(\mathbb{C}^{d_1}\otimes ... \otimes \mathbb{C}^{d_n})$,
\begin{equation}
        E_{D}^{ne}(\rho) = E_R^{\infty}(\rho).
\end{equation}
\end{proposition}
\begin{proof}
From Lemma \ref{singfractione} we find
\begin{equation} \label{sfne}
        F_{sep}(\rho^{\otimes n}; 2^{ny}) := \min_{\sigma \in 
        {\cal S}, b \in \mathbb{R}}\left[
        \tr (\rho^{\otimes n} - 2^{n b}
        \sigma)_+ + 2^{- (y - b)n}\right].
\end{equation}

Let us consider the asymptotic behavior of $F_{sep}(\rho^{\otimes n}, 
2^{n y})$. Take $y = E_{R}^{\infty}(\rho) + \epsilon$, for any 
$\epsilon > 0$. Then we can choose, for each $n$, 
$b = n(E_{R}^{\infty}(\rho) + \frac{\epsilon}{2})$, giving
\begin{equation*} 
        F_{sep}(\rho^{\otimes n}, 2^{ny}) \leq \min_{\sigma \in 
        {\cal S}} \left[ \tr(\rho^{\otimes n} - 2^{n(E_{\cal M}^{\infty}
        (\rho) + \frac{\epsilon}{2})}\sigma)_+\right] 
        + 2^{-n\frac{\epsilon}{2}}.
\end{equation*}
We then see from Eq. (\ref{compactmain}) that $\lim_{n \rightarrow 
\infty} F_{sep}(\rho^{\otimes n}, 2^{ny}) = 0$, from which follows 
that $E_D^{ne}(\rho) \leq E_{R}^{\infty}(\rho) + \epsilon$. As 
$\epsilon$ is arbitrary, we find $E_D^{ne}(\rho) \leq E_{R}^{\infty}
(\rho)$. 

Conversely, let us take $y = E_{R}^{\infty}(\rho) - \epsilon$, 
for any $\epsilon > 0$. The optimal $b$ for each $n$ has to satisfy 
$b_n \leq y$, otherwise $F_{sep}(\rho^{\otimes n}, 2^{ny})$ 
would be larger than one, which is not true. Therefore,
\begin{equation*} 
        F_{sep}(\rho^{\otimes n}, 2^{ny}) \geq \min_{\sigma \in 
        {\cal S}} \tr(\rho^{\otimes n} - 2^{n(E_{R}^{\infty}(\rho) 
        - \epsilon)}\sigma)_+,
\end{equation*}
which, by Eq. (\ref{compactmain}), tends to unity again. This then 
shows that $E_D^{ne}(\rho) \geq E_{R}^{\infty}(\rho) - \epsilon$. 
Again, as $\epsilon > 0$ is arbitrary, we find  $E_D^{ne}(\rho) 
\geq E_{R}^{\infty}(\rho)$.
\end{proof}

\vspace{0.3 cm}
The proof of the other half of Theorem \ref{maintheorem5} follows easily from Proposition \ref{disne} and the following Lemma. 

\begin{lemma} \label{almmonrelent}
If $\Lambda \in SEPP(\epsilon, {\cal H})$, then
\begin{equation} \label{1432}
        E_R(\Lambda(\rho)) \leq \log(1 + \epsilon) + E_R(\rho).
\end{equation}
\end{lemma}
\begin{proof}
Let $\sigma$ be an optimal separable state for $\rho$ in the relative 
entropy of entanglement. Then, if $\Lambda$ is a $\epsilon$-separability 
preserving map and $Z$ a state such that $\Lambda(\sigma) + \epsilon Z$ 
is separable,
\begin{eqnarray*}
        E_R(\rho) &=& S(\rho || \sigma) \nonumber \\ &\geq& 
        S( \Lambda(\rho) || \Lambda(\sigma) ) \nonumber \\
        &\geq& S( \Lambda(\rho) || \Lambda(\sigma) + \epsilon Z ) 
        \nonumber \\ 
        &=& S( \Lambda(\rho) || (\Lambda(\sigma) + 
        \epsilon Z)/(1 + \epsilon) ) - \log(1 + \epsilon) \nonumber \\ 
        &\geq& E_R(\Lambda(\rho)) - \log(1 + \epsilon), 
\end{eqnarray*}
The first inequality follows from the monotonicity of the relative 
entropy under trace preserving CP maps and the second inequality 
from the operator monotonicity of the $\log$.
\end{proof}
\vspace{0.3 cm}

Indeed, as any sequence of non-entangling maps is obviously 
asymptotically non-entangling, we have $E_D^{ane}(\rho) \geq 
E_D^{ne}(\rho) = E_R^{\infty}(\rho)$, where the last equality 
follows from Proposition \ref{disne}. To prove the converse 
inequality $E_D^{ane}(\rho) \leq E_R^{\infty}(\rho)$, we use 
Lemma \ref{almmonrelent}. Let $\Lambda_n \in 
SEPP(\epsilon_n)$ be an optimal sequence of maps for the 
distillable entanglement under asymptotically non-entangling 
maps in the sense that
\begin{equation*}
        \lim_{n \rightarrow \infty} || \Lambda_n(\rho^{\otimes n}) 
        - \phi_2^{\otimes k_n}||_1 = 0\;\;\;\;
        \;\;\;\; \lim_{n\rightarrow\infty} \epsilon_n = 0,
\end{equation*}
and
\begin{equation*}
\liminf_{n \rightarrow \infty} \frac{k_n}{n} = E_{D}^{ane}(\rho).
\end{equation*}
From Lemma \ref{almmonrelent},
\begin{eqnarray*}
       \frac{1}{n} E_R(\Lambda_n(\rho^{\otimes n}))
          &\leq&  \frac{1}{n} E_R(\rho^{\otimes n}) + \frac{1}{n} 
         \log(1 + \epsilon_n).
\end{eqnarray*}
Hence, as $\lim_{n \rightarrow \infty}\epsilon_n = 0$ and 
from the asymptotic continuity of relative entropy of entanglement,
\begin{eqnarray*}
        E_D^{ane}(\rho) &=& \liminf_{n \rightarrow \infty}
        \frac{1}{n}E_R(\Lambda_n(\rho^{\otimes n})) \\ 
        &\leq& \liminf_{n \rightarrow \infty} 
        \frac{1}{n}E_R(\rho^{\otimes n}) 
        + \liminf_{n \rightarrow \infty} \frac{1}{n} 
        \log(1 + \epsilon_n)\nonumber \\ 
        &=& E_R^{\infty}(\rho).
\end{eqnarray*}

\subsection{Proof Corollary \ref{corref}}

Finally, we can now easily establish Corollary \ref{corref}.
\vspace{0.2 cm}

\begin{proof} (Corollary \ref{corref})
We assume w.l.o.g. that $\sigma$ is entangled. Then, by Corollary ... of [], $E_R^{\infty}(\sigma) > 0$.

First, let us assume there is a sequence of quantum maps 
$\{ \Lambda_n \}_{n \in \mathbb{N}}$ satisfying the three conditions 
of the corollary. Then, 
\begin{eqnarray*}
        E_{R}^{\infty}(\sigma) &=& \lim_{n \rightarrow \infty} 
        \frac{1}{n}E_R(\Lambda_n(\rho^{\otimes n})) \\
        &\leq& \frac{1}{n}E_R(\rho^{\otimes n}) + 
        \frac{\log(1 + \epsilon_n)}{n} \\
        &=& E_R^{\infty}(\rho).
\end{eqnarray*}
The first equality follow from the asymptotic continuity of $E_R$ 
\cite{DH99} and the following inequality from Lemma \ref{almmonrelent}.

To show the other direction, let us assume that $E_R^{\infty}(\rho) 
\geq E_R^{\infty}(\sigma)$. As $E_R^{\infty}(\rho) = E_D^{ane}(\rho)$,
there is a sequence of maps $\{ \Lambda_n \}_{n \in \mathbb{N}}$, 
$\Lambda_n : {\cal D}((\mathbb{C}^{d_1}\otimes ... \otimes 
\mathbb{C}^{d_m})^{\otimes n}) \rightarrow {\cal D}((\mathbb{C}^{2} 
\otimes \mathbb{C}^{2})^{\otimes k_n})$, such that 
\begin{eqnarray*}
        \Lambda_n \in SEPP(\epsilon_n),  \hspace{1 cm} 
        \lim_{n \rightarrow \infty}\epsilon_n = 0, \nonumber 
\end{eqnarray*}
\begin{eqnarray} \label{tnbo1}  
        \lim_{n \rightarrow \infty} || \Lambda_n(\rho^{\otimes n}) 
        - \phi_2^{\otimes k_n} ||_1 = 0 \nonumber
\end{eqnarray}
and 
\begin{equation} \label{relentrho}
        \lim_{n \rightarrow \infty} \frac{k_n}{n} = E_R^{\infty}(\rho)
\end{equation}
Note we can always find a sequence for which the limit in Eq. (\ref{tnbo1})
exists by using the optimal sequence such that 
$\limsup_{n \rightarrow \infty} \frac{k_n}{n} = E_R^{\infty}(\rho)$ 
and increasing the value of the $k_n$'s which are not close to the 
limit value.

Moreover, as $E_R^{\infty}(\sigma) = E_C^{ane}(\sigma)$, there is 
another sequence of maps $\{ \Omega_n \}_{n \in \mathbb{N}}$, 
$\Omega_n : {\cal D}((\mathbb{C}^{2} \otimes 
\mathbb{C}^{2})^{\otimes k'_n}) \rightarrow {\cal D}
((\mathbb{C}^{d'_1}\otimes ... \otimes \mathbb{C}^{d'_{m'}})^{\otimes n})$, satisfying
\begin{eqnarray*}
        \Omega_n \in SEPP(\epsilon'_n),  \hspace{1 cm} \lim_{n \rightarrow \infty}\epsilon'_n = 0, \nonumber 
\end{eqnarray*}
\begin{eqnarray}        \label{tnbo2}
\lim_{n \rightarrow \infty} || \Omega_n(\phi_2^{\otimes k'_n}) - \sigma^{\otimes n} ||_1 = 0 \nonumber
\end{eqnarray}
and
\begin{equation} \label{relentsigma}
\lim_{n \rightarrow \infty} \frac{k'_n}{n} = E_R^{\infty}(\sigma).
\end{equation}

From Eqs. (\ref{relentrho}) and (\ref{relentsigma}) there is a 
sequence $\delta_{n_0}$ converging to zero when $n_0 \rightarrow 
\infty$ such that for every $n \geq n_0$,
\begin{equation*}
        k_n \geq (E_R^{\infty}(\rho) - \delta_{n_0}/2)n, \hspace{0.3 cm} k'_n \leq (E_R^{\infty}(\sigma) + \delta_{n_0}/2)n.
\end{equation*}
Then, for every $n \geq n_0$, $k_n \geq - \delta_{n_0}n + k'_n$. 
From Eq. (\ref{relentsigma}) we thus find that for sufficiently 
large $n \geq n_0$,
\begin{equation*}
        k_n =  k'_{n - o(n)} + r_n,
\end{equation*}
with $r_n$ a positive integer.

Let us now consider the sequence of maps $\{ \Omega_n \circ 
\tr_{1,...,r_n}\circ \Lambda_n \}_{n \in \mathbb{N}}$. From 
Eqs. (\ref{tnbo1}, \ref{tnbo2}) and the fact that the trace-norm
contracts under completely positive trace-preserving maps we find
\begin{eqnarray}
        \lim_{n \rightarrow \infty} || \Omega_{n - o(n)} \circ 
        \tr_{1,...,r_n} \circ \Lambda_n(\rho^{\otimes n}) - \sigma^{\otimes n - o(n)} ||_1  &\leq& \lim_{n \rightarrow \infty} || \Lambda_n(\rho^{\otimes n}) - \phi_2^{\otimes k_n} ||_1 \nonumber \\ &+& || \Omega_{n - o(n)}(\phi_2^{\otimes k'_{n - o(n)}}) - \sigma^{\otimes n - o(n)} ||_1 = 0. \nonumber
\end{eqnarray}
Moreover, from Lemma \ref{epsilonincrease} we see that for every separable state $\sigma$,
\begin{eqnarray*}
LR_G(\Omega_{n - o(n)} \circ \tr_{1,...,r_n} \circ \Lambda_n(\sigma)) &\leq& LR_G(\Lambda_n(\sigma)) + \log(1 + \epsilon'_n) \\ &\leq& \log(1 + \epsilon_n) + \log(1 + \epsilon'_n),
\end{eqnarray*}
where we used $\Omega_{n - o(n)} \circ \tr_{1,...,r_n} \in SEPP(\epsilon'_n)$ and $\Lambda_n \in SEPP(\epsilon_{n - o(n)})$. Hence, $\Omega_{n - o(n)} \circ \tr_{1,...,o(n)} \circ \Lambda_n \in SEPP(\epsilon_n + \epsilon'_n + \epsilon_n \epsilon'_n)$.
\end{proof}

\textbf{Remark:} The structure of the proof can be applied to other 
situations apart from entanglement conversion. First, as discussed in Ref. \cite{BP08}, Theorem \ref{maintheorem} holds true not only to 
discrimination of an entangled state from a sequence of separable 
states, but also to the discrimination of any i.i.d. quantum state 
from a sequence of states belonging to sets ${\cal M}_n$, satisfying 
five certain properties (see Ref. \cite{BP08} for details). In 
addition to Theorem \ref{maintheorem}, the only particular property 
of entangled states that we used is that (i) $\Phi(K)$ is entangled; 
(ii) the largest fidelity of $\Phi(K)$ with a separable state is 
$1/K$; an (iii) isotropic states (convex combinations of $\Phi(K)$ 
and its orthogonal state) are separable iff the weight of $\Phi(K)$ 
is smaller than $1/K$. Therefore, Theorem \ref{maintheorem5} is true 
in other settings, as long as the properties mentioned before remain 
true if we change the set of separable states for another one. For 
example, we can find similar conclusions for a conversion theory of 
states with a non-positive partial transpose, where PPT states are 
considered in the place of separable states.

\section{How Much Entanglement Must and Can be Generated?}  
\label{mustcan}

We are now in position to understand the choice of the global 
robustness as the measure to quantify the amount of entanglement 
generated. The reason that we need to allow some entanglement to 
be generated is that we relate the entanglement cost to the 
regularized relative entropy of entanglement by using the connection 
of the latter to the asymptotic global robustness. The amount of 
entanglement generated is then due to the fact that the optimal 
mixing state in the global robustness might be entangled. Before 
we analyse more carefully if we indeed need to allow for some 
entanglement to be generated, let us analyse if we can quantify 
it by some other measure, instead of the global robustness.

Suppose we required alternatively only that 
\begin{equation} \label{tnapp}
        \lim_{n \rightarrow \infty} \max_{\sigma \in {\cal S}} 
        \min_{\pi \in {\cal S}} || \Lambda_n(\sigma) - \pi ||_1 = 0,
\end{equation}
instead of $\lim_{n \rightarrow \infty}\max_{\sigma \in {\cal S}} 
R_G(\Lambda_n(\sigma)) = 0$. Then the achievability part in 
Proposition \ref{disne} would still hold, as we use operations which do 
not generate any entanglement, i.e. they map separable states to 
separable states.

However this is not sufficient. We still have to make sure that 
the cost is larger than the distillation function, which should 
be finite. It is easy to see that Eq. (\ref{tnapp}) ensures that 
both the distillation and cost functions are zero for separable 
states. It turns out however that the distillable entanglement 
is infinite for every entangled state! We hence have a bizarre 
situation in which even though entanglement cannot be created 
for free, it can be amplified to the extreme whenever present, 
no matter in what amount. The key to see this is to consider 
the analogue of $F_{sep}$, given by Eq. (\ref{singletfractionoriginal}), when we 
only require that the map satisfies Eq. \ref{tnapp}. Following 
the proof of Lemma \ref{singfractione} we can easily see that 
the singlet-fraction under maps $\Lambda$ satisfying 
\begin{equation*} 
        \max_{\sigma \in {\cal S}} \min_{\pi \in {\cal S}} || 
        \Lambda(\sigma) - \pi ||_1 \leq \epsilon
\end{equation*}
is given by
\begin{equation*} 
        F_{sep}(\rho; K ; \epsilon) = \min_{\sigma \in \text{cone}
        ({\cal S})}\left[\tr (\rho - \sigma)_+ + \tr(\sigma)(\frac{1}{K} 
        + \epsilon)\right],
\end{equation*}
which for $\rho^{\otimes n}$ can be rewritten as
\begin{equation*} 
        F_{sep}(\rho^{\otimes n}; 2^{ny}; \epsilon_n) = \min_{\sigma 
        \in {\cal S}, b \in \mathbb{R}} 
        \left[ \tr (\rho^{\otimes n} - 2^{bn}\sigma)_+ 
        + 2^{-(y - b)n} + 2^{-( (\log(1/\epsilon_n)/n)
        - b)n}\right].
\end{equation*}
It is clear that the optimal $b$ must be such that 
$b < \min(y, \log(1/\epsilon_n)/n)$, as otherwise 
$F_{sep}(\rho^{\otimes n}; 2^{ny}; \epsilon)$ would be larger 
than unity. Therefore, if $y > \log(1/\epsilon_n)/n$, 
\begin{equation*}
        F_{sep}(\rho^{\otimes n}; 2^{ny}; \epsilon_n) \geq 
        \min_{\sigma \in {\cal S}}  \tr (\rho^{\otimes n} - 
        \epsilon_n^{-1}\sigma)_+.
\end{equation*} 
By Theorem \ref{maintheorem}, $F_{sep}(\rho^{\otimes n}; 2^{ny}; 
\epsilon_n)$ approaches unity for every $y$, as long as $\epsilon_n$ 
goes to zero slower than $2^{- n E_R^{\infty}(\rho)}$, which 
implies that the associated distillable entanglement is unbounded. 
Note that the same happens if we use \textit{any} asymptotically 
continuous measure to bound the amount of entanglement generated. Here we denote a measure $E$ is asymptotically continuous if for all states $\rho, \sigma \in {\cal D}({\cal H})$, $|E(\rho) - E(\sigma)| \leq \log({\text{dim}}({\cal H})) f(|| \rho_n - \sigma_n ||_1)$, for a real valued function $f$ independent of ${\text{dim}}({\cal H})$ such that $\lim_{x \rightarrow \infty}f(x) = 0$.

If instead we require that
\begin{equation*}
        \max_{\sigma \in {\cal S}} \min_{\pi \in {\cal S}} || 
        \Lambda(\sigma) - \pi ||_1 \leq \epsilon/ \dim({\cal H}),
\end{equation*}
or even that 
\begin{equation*}
        \max_{\sigma \in {\cal S}} \min_{\pi \in {\cal S}} || 
        \Lambda(\sigma) - \pi ||_{\infty} \leq \epsilon/ 
        \dim({\cal H}),
\end{equation*}
then we would find that the associated $\epsilon$-singlet-fraction 
would satisfy
\begin{equation*} 
        \tilde{F}_{sep}(\rho; K ; \epsilon) = \min_{\sigma \in 
        \text{cone}({\cal S})} \left[ \tr (\rho - \sigma)_+ + 
        \tr(\sigma)\frac{1 + \epsilon}{K}\right].
\end{equation*}
In this case it is easy to see that the distillable entanglement 
would be bounded and we would recover a sensible situation. It is 
interesting and rather mysterious to the authors that although it 
seems that some entanglement must be generated to have reversibility, 
only very little can actually be afforded before the theory becomes 
trivial.

For analysing the necessity of generating some entanglement for 
reversibility, we consider the following variant of $R_G$ \cite{VT99}:
\begin{equation*}
        R(\rho) = \min_{\sigma \in {\cal S}, s \in \mathbb{R}} 
        \left[s : \frac{\rho + s \sigma}{1 + s} \in {\cal S}\right],
\end{equation*}
and its log version $LR(\rho) := \log(1 + R(\rho))$. Then, in analogy 
to $LG$, we define 
\begin{equation*} 
        LH(\rho) := \inf_{\{ \epsilon_n\}} \left 
        \{ \limsup_{n \rightarrow \infty} \frac{1}{n} LR^{\epsilon_n}
        (\rho^{\otimes n}) : \lim_{n \rightarrow \infty} \epsilon_n 
        = 0 \right \}, 
\end{equation*}
where  
\begin{equation} 
        LR^{\epsilon}(\rho) := \min_{\tilde{\rho} \in B_{\epsilon}(\rho)} 
        LR(\tilde{\rho}), 
\end{equation}
with $B_{\epsilon}(\rho) := \{ \tilde{\rho} \in {\cal D}({\cal H}) : 
|| \rho- \tilde{\rho} ||_1 \leq \epsilon \}$. Following the proof of 
Proposition \ref{propcost} 
it is straightforward to show that the entanglement cost under 
strictly non-entangling maps is given by $LH$. Therefore, the 
question whether we must allow the generation of some entanglement 
in order to have a reversible theory reduces to the question whether 
the two robustness measures $LG$ and $LH$ become the same quantity 
after smoothing and regularization.

\section{Connection to the Axiomatic Formulation of the Second Law of Thermodynamics} \label{cafslt}

In this section we comment on the similarities and differences of 
entanglement manipulation under asymptotically non-entangling 
operations and the axiomatic approach of Giles \cite{Gil64} and 
more particularly of Lieb and Yngvason \cite{LY99} for the second 
law of thermodynamics. It must be emphasized that we are comparing 
the two theories only on an abstract level. Indeed, although the 
two theories deals with completely different resources (entanglement 
and order) and have also a distinct range of applicability, we find 
that the structural form of them both is actually very much 
related.

Let us start by briefly recalling the axioms used in Ref. \cite{LY99} in order to derive the second law. Their starting point is 
the definition of a system as a collection of points called state space 
and denoted by $\Gamma$. The individual points of a state space are 
the states of the system. The composition of two state spaces 
$\Gamma_1$ and $\Gamma_2$ is given by their Cartesian product. 
Furthermore, the scaled copies of a given system are defined as 
follows: if $t > 0$ is some fixed number, the state space 
$\Gamma^{(t)}$ consists of points denoted by $t X$ with $X \in \Gamma$. 
Finally, a preorder $\prec$ on the state space satisfying the 
following axioms is assumed:
\begin{enumerate}
        \item $X \prec X$.
        \item $X \prec Y$ and $Y \prec Z$ implies $X \prec Z$.
        \item If $X \prec Y$, then $t X \prec tY$ for all $t > 0$.
        \item $X \prec (t X, (1 - t)X)$ and $(t X, (1 - t)X) \prec X$ for all $0 \leq y \leq 1$.
        \item If, for some pair of states, $X$ and $Y$,
        \begin{equation}
        (X, \epsilon Z_0) \prec (Y, \epsilon Z_1)
\end{equation}
holds for a sequence of $\epsilon$'s tending to zero and some states 
$Z_0$, $Z_1$, then $X \prec Y$.
\item $X \prec X'$ and $Y \prec Y'$ implies $(X, Y) \prec (X', Y')$.
\end{enumerate}
It was then shown that these axioms, together with the 
comparison hypothesis, which states that 
\begin{center}
\textbf{Comparison Hypothesis:} for any two states $X$ 
and $Y$ in the same state space $\Gamma$, either $X \prec Y$ or $Y \prec X$,
\end{center}
are sufficient to prove the existence of a single valued entropy 
function completely determining the order induced by the relation 
$\prec$. More precisely, one also need the vality of the comparison hypothesis for all two-fold scaled products $(1 - t)\Gamma \times t \Gamma$ of the state space $\Gamma$. 

In the context of entanglement transformations, we interpret the 
relation $\rho \prec \sigma$ as the possibility of asymptotically 
transforming $\rho$ into $\sigma$ by asymptotically non-entangling
maps. Then, the composite state $(\rho, \sigma)$ is nothing but 
the tensor product $\rho \otimes \sigma$. Moreover, $t \rho$ takes 
the form of $\rho^{\otimes t}$. Then $\rho^{\otimes t} \prec \sigma$
expresses the fact that asymptotically $t$ copies of $\rho$ can be 
transformed into one of $\sigma$. More concretely, we say that 
\begin{equation*}
        \rho^{\otimes t} \prec \sigma^{\otimes q},
\end{equation*}
for positive real numbers $t, q$ if there is a
sequence of integers $n_t, n_q$ and of $SEPP(\epsilon_n)$ maps
$\Lambda_n$ such that
\begin{equation*}
        \lim_{n \rightarrow \infty} || \Lambda_{n}(\rho^{\otimes n_t}) 
        - \sigma^{\otimes n_q - o(n)}||_1 = 0,
\end{equation*}
\begin{equation*}
        \lim_{n \rightarrow \infty} \epsilon_n = 0, \hspace{0.3 cm} 
        \lim_{n \rightarrow \infty} \frac{n_t}{n} = t, \hspace{0.3 cm} 
        \text{and}\hspace{0.3 cm} \lim_{n \rightarrow \infty} 
        \frac{n_q}{n} = q.
\end{equation*}
With this definition it is straightforward to observe that properties 
$1$, $3$, and $4$ hold true for entanglement manipulation under 
asymptotically non-entangling maps. Property 2 can be shown to hold, 
in turn, by noticing that, from Lemma \ref{epsilonincrease}, if 
$\Lambda \in SEPP(\epsilon)$ and $\Omega \in SEPP(\delta)$, then 
$\Lambda \circ \Omega \in SEPP(\epsilon + \delta + \delta \epsilon)$. 
Therefore the composition of two asymptotically non-entangling maps 
is again asymptotically non-entangling. That property 5 is also true 
is proven in the following lemma.

\begin{lemma} \label{lemmaoasifhksdhfkjsfd}
If for two states $\rho$ and $\sigma$,
\begin{equation} \label{axi1}  
        \rho \otimes \pi_1^{\otimes \epsilon} \prec 
        \sigma \otimes \pi_2^{\otimes \epsilon},
\end{equation} 
holds for a sequence of $\epsilon$'s tending to zero and two states 
$\pi_0$, $\pi_1$, then $\rho \prec \sigma$.
\end{lemma}
\begin{proof}
Eq. (\ref{axi1}) means that for every $\epsilon > 0$ there is a 
sequence of maps $\Lambda_n \in SEPP(\epsilon_n)$ such that
\begin{equation*}
        \lim_{n \rightarrow \infty} || \Lambda_{n}(\rho^{\otimes n} 
        \otimes \pi_1^{\otimes n_{\epsilon}}) - \sigma^{\otimes n 
        - o(n)} \otimes \pi_2^{\otimes n'_{\epsilon} - o(n)}||_1 
        = \lim_{n \rightarrow \infty} \delta_n = 0,
\end{equation*}
\begin{equation*}
        \lim_{n \rightarrow \infty} \epsilon_n = 0, \hspace{0.3 cm} 
        \lim_{n \rightarrow \infty} \frac{n_{\epsilon}}{n} = \epsilon, 
        \hspace{0.3 cm} \text{and} \hspace{0.3 cm} \lim_{n \rightarrow 
        \infty} \frac{n'_{\epsilon}}{n} = \epsilon.
\end{equation*}
We have
\begin{eqnarray*}
        \frac{1}{n}E_R(\rho^{\otimes n}) + \frac{1}{n}E_R(
        \pi_1^{\otimes n_{\epsilon}}) &\geq& \frac{1}{n} E_R(
        \rho^{\otimes n} \otimes \pi_1^{\otimes n_{\epsilon}}) 
        \nonumber \\ 
        &\geq& \frac{1}{n}E_R(\Lambda_{n}(\rho^{\otimes n} \otimes 
        \pi_1^{\otimes n_{\epsilon}})) - \frac{\log(1 + \epsilon_n)}{n} 
        \nonumber \\ 
        &\geq& \frac{1}{n}E_R(\sigma^{\otimes n - o(n)} \otimes 
        \pi_2^{\otimes n'_{\epsilon} - o(n)}) - f(\delta_{\epsilon}) - 
        \frac{\log(1 + \epsilon_n)}{n} \nonumber \\
        &\geq& \frac{1}{n}E_R(\sigma^{\otimes n - o(n)}) - 
        f(\delta_{\epsilon}) - \frac{\log(1 + \epsilon_n)}{n}, 
\end{eqnarray*}
where $f: \mathbb{R} \rightarrow \mathbb{R}$ is such that 
$\lim_{x \rightarrow 0}f(x) = 0$. The first inequality follows from 
the subadditivity of $E_R$, the second from Lemma \ref{almmonrelent}, 
the third from the asymptotic continuity of $E_R$, and the last from 
the monotonicity of $E_R$ under the partial trace. 

As $E_R(\pi_2) \leq \log(\dim({\cal H}))$, where ${\cal H}$ is the 
Hilbert space in which $\pi_2$ acts on, we find 
\begin{equation*}
        \frac{1}{n}E_R(\rho^{\otimes n}) \geq \frac{1}{n}E_R(
        \sigma^{\otimes n - o(n)}) - f(\delta_{\epsilon}) - 
        \frac{\log(1 + \epsilon_n)}{n} - \frac{n_{\epsilon}}{n}
        \log(\dim({\cal H})).
\end{equation*}
Taking the limit $n \rightarrow \infty$,
\begin{equation*}
        E_R^{\infty}(\rho) \geq E_R^{\infty}(\sigma)  - \epsilon.
\end{equation*}
Taking $\epsilon \rightarrow 0$ we find that $E_R^{\infty}(\rho) 
\geq E_R^{\infty}(\sigma)$. The Lemma then follows from Corollary 
\ref{corref}.
\end{proof}

The Comparison Hypothesis, in turn, follows from Corollary 
\ref{corref}: it expresses the total order 
induced by the regularized relative entropy of entanglement. 

We do not know if the theory we are considering for entanglement 
satisfy axiom 6. This is fundamentally linked to the possibility 
of having \textit{entanglement catalysis} \cite{JP99} under 
asymptotically non-entangling transformations. One can prove the following 
simple lemma. 

\begin{lemma}
For entanglement transformations under asymptotically non-entangling maps, axiom 6 is equivalent to
\begin{equation} \label{noncatalysis}
\text{If there is a $\pi$ such that} \hspace{0.2 cm} \rho \otimes \pi \prec \sigma \otimes \pi, \hspace{0.2 cm} \text{then} \hspace{0.2 cm} \rho \prec \sigma.
\end{equation}
\end{lemma}

\begin{proof}
In Theorem 2.1 of Ref. \cite{LY99} is was shown that axiom 1-6 implies Eq. (\ref{noncatalysis}). Since entanglement manipulations under asymptotically non-entangling maps satisfies axioms 1-5, we find one direction of the equivalence. 

To prove the converse, assume Eq. (\ref{noncatalysis}) holds true. Following \cite{LY99}, we use $X \prec\prec Y$ to denote the situation in which $X \prec Y$, but the reverse transformation is impossible. We claim that Eq. (\ref{noncatalysis}) implies 
\begin{equation} \label{auxauxauxaux}
\rho  \prec \prec \sigma \Rightarrow \rho \otimes \pi \prec \sigma \otimes \pi \hspace{0.4 cm} \forall \hspace{0.1 cm} \pi.
\end{equation}

Before we prove this implication, let us show how we can use Eq. (\ref{auxauxauxaux}) to get the result. Let $\rho_1, \rho_2, \sigma_2, \sigma_2$ be such that $\rho_1 \prec \sigma_1$ and 
$\rho_2 \prec \sigma_2$. Then, by Corollary \ref{corref} and the weak 
additivity of $E_R^{\infty}$, we find 
$\rho_1^{\otimes 1 + \epsilon} \prec \prec \sigma_1$ and $\rho_2^{\otimes 1 + \epsilon} 
\prec \prec \sigma_2$, for every $\epsilon > 0$. Then, applying Eq. (\ref{auxauxauxaux}) twice,
\begin{equation}
\rho_1^{\otimes 1 + \epsilon} \otimes \rho_2^{\otimes 1 + \epsilon} \prec \sigma_1 \otimes \rho_2^{\otimes 1 + \epsilon} \prec \sigma_1 \otimes \sigma_2.
\end{equation}
The result of the lemma follows from Lemma \ref{lemmaoasifhksdhfkjsfd} and the fact that $\epsilon > 0$ is arbitrary. 

Let us now turn to the derivation of Eq. (\ref{auxauxauxaux}). We actually show that the negation of Eq. (\ref{auxauxauxaux}) implies the negation of Eq. (\ref{noncatalysis}). Indeed the former reads 
\begin{equation} \label{ahsdklahfdds}
\text{NOT} (\ref{auxauxauxaux}) : \text{there is a triple} \hspace{0.2 cm} \rho, \sigma, \pi \hspace{0.2 cm} \text{such that} \hspace{0.2 cm} \rho  \prec \prec \sigma \hspace{0.2 cm} \text{and} \hspace{0.2 cm} \text{NOT} \hspace{0.1 cm} \rho \otimes \pi \prec \sigma \otimes \pi.
\end{equation}
The total order established in Corollary \ref{corref} shows that impossibility of the transformation $\rho \otimes \pi \prec \sigma \otimes \pi$ is equivalent to $\sigma \otimes \pi \prec \prec \rho \otimes \pi$. Then we can rewrite Eq. (\ref{ahsdklahfdds}) as
\begin{equation*} 
\text{NOT} (\ref{auxauxauxaux}) : \text{there is a triple} \hspace{0.2 cm} \rho, \sigma, \pi \hspace{0.2 cm} \text{such that} \hspace{0.2 cm} \rho  \prec \prec \sigma \hspace{0.2 cm} \text{and} \hspace{0.2 cm} \sigma \otimes \pi \prec \prec \rho \otimes \pi.
\end{equation*}
To make the identification simpler let us make the relabeling $\rho \leftrightarrow \sigma$ in the equation above to get
\begin{equation} \label{finalofthislemmabeforethelast}
\text{NOT} (\ref{auxauxauxaux}) : \text{there is a triple} \hspace{0.2 cm} \rho, \sigma, \pi \hspace{0.2 cm} \text{such that} \hspace{0.2 cm} \sigma  \prec \prec \rho \hspace{0.2 cm} \text{and} \hspace{0.2 cm} \rho \otimes \pi \prec \prec \sigma \otimes \pi.
\end{equation}

The negation of Eq. (\ref{noncatalysis}), in turn, is the following
\begin{equation*} 
\text{NOT} (\ref{noncatalysis}) : \text{there is a triple} \hspace{0.2 cm} \rho, \sigma, \pi \hspace{0.2 cm} \text{such that} \hspace{0.2 cm} \rho \otimes \pi \prec \sigma \otimes \pi \hspace{0.2 cm} \text{and} \hspace{0.2 cm} \text{NOT} \hspace{0.1 cm} \rho \prec \sigma.
\end{equation*}
From Corollary \ref{corref}, once more, we have that the negation of $\rho \prec \sigma$ is equivalent to $\sigma \prec \prec \rho$. Thus
\begin{equation} \label{finalofthislemmabeforethelast2}
\text{NOT} (\ref{noncatalysis}) : \text{there is a triple} \hspace{0.2 cm} \rho, \sigma, \pi \hspace{0.2 cm} \text{such that} \hspace{0.2 cm} \rho \otimes \pi \prec \sigma \otimes \pi \hspace{0.2 cm} \text{and} \hspace{0.2 cm} \sigma \prec \prec \rho.
\end{equation}
It is now clear that Eq. (\ref{finalofthislemmabeforethelast}) implies Eq. (\ref{finalofthislemmabeforethelast2}).
\end{proof}

We can link such a possibility of catalysis in the bipartite case to an important 
open problem in entanglement theory, the full additivity of the 
regularized relative entropy of entanglement. In turn, the latter 
was shown in Ref. \cite{BHPV07} to be equivalent to the full 
monotonicity under \textit{LOCC} of $E_R^{\infty}$. 

\begin{lemma}
The regularized relative entropy of entanglement is fully additive for bipartite states, i.e. for every two states 
$\rho \in {\cal D}(\mathbb{C}^{d_1}\otimes \mathbb{C}^{d_2})$ and $\pi \in {\cal D}(\mathbb{C}^{d'_1}\otimes \mathbb{C}^{d'_2})$,
\begin{equation} \label{relentadd}
E_{R}^{\infty}(\rho \otimes \pi) = E_{R}^{\infty}(\rho) + E_{R}^{\infty}(\pi),
\end{equation}
if, and only if, there is no catalysis for entanglement manipulation under 
asymptotically non-entangling maps.
\end{lemma}
\begin{proof}
If Eq. (\ref{relentadd}) holds true and $\rho \otimes \pi \prec \sigma \otimes \pi$, then
\begin{equation*}
E_{R}^{\infty}(\rho) + E_{R}^{\infty}(\pi) = E_{R}^{\infty}(\rho \otimes \pi) \geq E_{R}^{\infty}(\sigma \otimes \pi) = E_{R}^{\infty}(\sigma) + E_{R}^{\infty}(\pi),
\end{equation*}
and thus, as $E_{R}^{\infty}(\rho) \geq E_{R}^{\infty}(\sigma)$, we find from Corollary \ref{corref} that $\rho \prec \sigma$. 

Conversely, assume that there is no catalysis. Then from the discussion above we find that axiom 6 holds true. For every bipartite pure state $\ket{\psi}$, the regularized relative entropy of entanglement is equal to the von Neumann entropy of the reduced density matrix $S(\psi_A)$. It hence follows that for every bipartite state $\rho$, there is a bipartite pure state $\ket{\psi}$ such that $E_R^{\infty}(\rho) = E_R^{\infty}(\psi)$. 

Let $\ket{\psi}$ and $\ket{\phi}$ be such that $E_R^{\infty}(\rho) = E_R^{\infty}(\psi)$ and $E_R^{\infty}(\pi) = E_R^{\infty}(\phi)$. From Corollary \ref{corref} we have $\rho \prec \psi$, $\pi \prec \phi$ and vice versa. Then, by axiom 6 we find that $\rho \otimes \pi \prec \psi \otimes \phi$ and $\psi \otimes \phi \prec \rho \otimes \pi$, from which we find, once more from Corollary \ref{corref}, that $E_R^{\infty}(\rho \otimes \pi) = E_R^{\infty}(\psi \otimes \phi)$. The lemma is a consequence of the additivity of $E_R^{\infty}$ on two pure states (which follows from the fact that for pure states the measure is equal to the entropy of entanglement). 
\end{proof}

It is an open question if we can extend the lemma to the multipartite 
setting. The difficulty in this case is that we do not have a simple 
formula for $E_R^{\infty}$ of pure states and hence do not know if the 
measure is additive for two multipartite pure states.

{\em Acknowledgements --} We gratefully acknowledge Koenraad Audenaert, 
Jens Eisert, Andrzej Grudka, Micha\l{} Horodecki, Ryszard Horodecki, Shashank 
Virmani, Reinhard Werner, Andreas Winter, and the participants in the 2009 McGill-Bellairs workshop for many interesting discussions and useful 
correspondences. This work 
is part of the QIP-IRC supported by EPSRC (GR/S82176/0) as well as the 
Integrated Project Qubit Applications (QAP) supported by the IST 
directorate as Contract Number 015848 and was supported by the 
Brazilian agency Conselho Nacional de Desenvolvimento Cient\'ifico 
e Tecnol\'ogico (CNPq), an EPSRC Postdoctoral Fellowship for 
Theoretical Physics and a Royal Society Wolfson Research Merit Award.

\end{document}